\documentclass[12pt]{article}
\usepackage{epsfig}

\usepackage{amssymb}
\usepackage{amsmath}
\usepackage{amsfonts}

\usepackage{amsthm}

\theoremstyle{theorem}

\theoremstyle{definition}

\def\bp{\begin{proof}}
\def\ep{\end{proof}}

\def\be{\begin{equation}}
\def\ee{\end{equation}}
\def\ba{\begin{array}{c}}
\def\ea{\end{array}}

\def\ben{\[}
\def\een{\]}

\newcommand{\bea}{\begin{eqnarray}}
\newcommand{\eea}{\end{eqnarray}}

\newcommand{\bbr}{\br\!\br}
\newcommand{\kkt}{\kt\!\kt}

\newcommand{\kt}{\rangle}
\newcommand{\br}{\langle}
\begin{document}

\titlepage

\vspace{.35cm}

 \begin{center}{\Large \bf

Discrete quantum square well of the first kind.

  }\end{center}

\vspace{10mm}

 \begin{center}

 {\bf Miloslav Znojil}

 \vspace{3mm}
Nuclear Physics Institute ASCR,

250 68 \v{R}e\v{z}, Czech Republic

{e-mail: znojil@ujf.cas.cz}

\vspace{3mm}

%
%\today, chebypol.tex

\end{center}

\vspace{5mm}

%\newpage

\section*{Abstract}

A new exactly solvable cryptohermitian quantum chain model is
proposed and analyzed. Its discrete-square-well-like Hamiltonian
with the real spectrum possesses a manifestly non-Hermitian form. It
is only made self-adjoint by the constructive transition to an {\em
ad hoc} Hilbert space. Such a space (i.e., the closed form of its
inner product, i.e., the ``metric" $\Theta$) varies with an $N-$plet
of optional parameters. The simplicity of our model enables one to
obtain the complete family of these physics-determining metrics
$\Theta$ in a user-friendly band-matrix closed form.

%\newpage

\section{Introduction \label{I} }

One of the most remarkable features of quantum mechanics may be seen
in the robust nature of its ``first principles" and probabilistic
interpretation. These principles practically did not change during
the last cca eighty years. In a sharp contrast, the production of
the innovative applications of the theory does not seem to have ever
slowed down.

In particular, twenty years ago, Scholtz et al \cite{Geyer}
emphasized that the {\em dynamical} content of the phenomenological
models of microscopic systems may be encoded not only in the
Hamiltonians and other observables but also, equally efficiently, in
metric operators $\Theta$ defining certain nontrivial,
``sophisticated" representations ${\cal H}^{(S)}$ of the physical
Hilbert space of the bound states of the quantum  system in
question\footnote{in {\it loc. cit.} the authors decided to denote
the Hilbert-space metric by the symbol $T$; we shall prefer the use
of the ``Greek translation" of this letter - see also \cite{SIGMA}
for explanation}.

It is worth noticing that in {\it loc. cit.} the authors had in
mind, first of all, the physical systems as ``traditional" as the
heavy atomic nuclei. During the recent renaissance of the theory the
use and  appeal of nontrivial metrics $\Theta^{(S)}\neq I$ has
significantly been extended. In particular, the recent reviews
\cite{Carl,Dorey,ali} of the so called ${\cal PT}-$symmetric quantum
mechanics may be recalled as describing, in detail, a number of the
less traditional explicit models, with the applicability ranging
from the very pragmatic phenomenological fits of spectra up to the
ambitious theoretical proposals throwing new light on various old
questions (the most recent proceedings \cite{proc} may be also
recalled as a recommended introductory reading).

In the particular ${\cal PT}-$symmetric theoretical framework one
selects just a very specific subclass of eligible metrics
$\Theta^{(S)}_{\cal PT}$ which prove expressible as products of
parity ${\cal P}$ and charge ${\cal C}$. In practice, unfortunately,
the simplicity of such a recipe appeared to be accompanied by its
conceptual weakness and non-applicability in the scattering regime
\cite{Jones}. This strongly motivated a subsequent return to the
study of the other models characterized by the potentially broader
variability of the Hilbert-space metric \cite{scatt,fund}.

Unfortunately, the latter return to the less straightforward
model-building recipes has been accompanied by the perceivable
growth of technical difficulties \cite{Jonesdva}. For this reason,
the attention of many authors has been re-attracted to the quantum
models in which the use of a nontrivial metric $\Theta^{(S)}\neq
\Theta^{(S)}_{\cal PT}$ gets combined with very schematic and
``user-friendly" Hamiltonians $H$. {\it Pars pro toto} let us
recollect one of the most elementary illustrative examples provided
by the square-well Hamiltonian $H$ of Ref.~\cite{sqw} accompanied by
the nontrivial metrics $\Theta^{(S)}$ proposed by Mostafazadeh and
Batal \cite{Batal}.

The latter authors emphasized that the physical content of the
quantum models described by the {\em pairs} of the operators
$(H,\Theta)$ remains entirely standard and compatible not only with
the very general recipe and formalism of Ref.~\cite{Geyer} but also,
at least implicitly, with any current textbook on quantum mechanics.
In particular, the authors constructed the related position and
momentum operators and explained the existence of the conserved
probability density as well as the feasibility of the more or less
standard backward transition to the classical, non-quantum limit.

In what follows we intend to describe a new square-well-related
quantum model. Although we shall mainly pay attention to its formal
aspects, physics-motivated readers may equally well reveal a strong
phenomenological appeal of our present model in its very  close
formal as well as qualitative link to the above-mentioned
square-well system. Moreover, in the light of the recent literature
the independent and purely experimental appeal of the similar models
has recently been found even in {\em non-quantum} physics
\cite{optical}.

Our message will be organized as follows. In section \ref{Ia} we
shall review a few older relevant results. Subsequently, in section
\ref{III} we shall restrict our attention to the new, extremely
elementary Chebyshev-lattice Hamiltonian. We shall describe the
properties of this quantum model in detail. In section \ref{IV} we
shall finally pay attention to the parametric-dependence of the
associated ``sophisticated" physical Hilbert space(s).

All of these results will indicate that even when the Hamiltonian
matrix itself remains comparatively elementary, the quantum dynamics
of the system may remain rich in a way mediated by the variability
of the``sophisticated" physical Hilbert space. The existence of
certain Hilbert-space-induced boundaries of the observability of the
quantum system in question will be pointed out. Their shape will be
shown to vary with our choice of the optional parameters determining
the (ambiguous) Hilbert space. This and other observations will be
explained and summarized in section \ref{V}.

\section{Framework: Quantum chain models \label{Ia} }

\subsection{Discrete quantum square wells \label{uII} }

It is well known \cite{disqw} that the discrete square-well
Schr\"{o}dinger equation
 \be
 {H}^{[U]}\, |\psi^{[U]}_n\kt= E^{[U]}_n\,
|\psi^{[U]}_n\kt\,, \ \ \ \ \ n = 0, 1, \ldots, N-1\,
 \label{SEwodelU}
  \ee
with the real, $N$ by $N$ Hamiltonian matrix
 \be
 H^{[U]}=\left[
\begin {array}{ccccc}
0&1&0&\ldots&0\\
{}1&0&1&\ddots&\vdots\\
{}0&1&\ddots&\ddots&0\\
{}\vdots&\ddots&\ddots&0&1\\
{}0&\ldots&0&1&0
\end {array} \right]\,
 \label{sqwodel}
 \ee
is exactly solvable in terms of the Chebyshev's polynomials of the
second kind\footnote{our denotation of these polynomials is borrowed
from MAPLE \cite{Maple}; for our present purposes it proves much
better suited than their more common subscripted
symbols~\cite{RyzhikStegun}.},
 \be
 |\psi_n^{[U]}\kt=\left[
 \begin {array}{c}
 U(0,  x_n)\\
 U(1,  x_n)\\
 \vdots\\
 U(N-1,  x_n)
 \end {array} \right]\,.
 \label{eigvecU}
 \ee
On this background, let us call this model, for our present
purposes, the discrete quantum square well of the second kind. For
this model it is worth noticing that the related bound-state
spectrum of energies $E_n^{[U]}=2x_n$ is all available in closed
form,
 \be
 E^{[U]}_n=2\,\cos
 \frac{(n+1)\pi}{N+1}\,, \ \ \ \ \ n = 0, 1, \ldots, N-1\,\,.
 \label{eigvalU}
 \ee
This ``square-well model of the second kind" can immediately be
identified with the discrete version of the standard
Schr\"{o}dinger's differential-equation square-well problem
\cite{Fluegge}. Hence, its large$-N$ limit remains well understood
and solvable.

In our present letter we intend to introduce a similar ``square-well
model of the first kind" obtained via a replacement of the elements
of the quantum state vectors (\ref{eigvecU}) by the equally friendly
Chebyshev polynomials of the first kind,
 \be
 |\psi_n^{[T]}\kt=\left[
 \begin {array}{c}
 T(0,  x_n)\\
 T(1,  x_n)\\
 \vdots\\
 T(N-1, x_n)
 \end {array} \right]\,.
 \label{eigvecT}
 \ee
It is easy to show that with $E_n^{[T]}=2x_n$  such an ansatz will
be compatible with the alternative square-well-type Schr\"{o}dinger
equation
 \be
 {H}^{[T]}\, |\psi^{[T]}_n\kt= E^{[T]}_n\,
|\psi^{[T]}_n\kt\,, \ \ \ \ \ n = 0, 1, \ldots, N-1\,
 \label{SEwodelT}
  \ee
in which another real though manifestly non-Hermitian
Hamiltonian-type matrix is being used,
 \be
 H^{[T]}=\left[
\begin {array}{ccccc}
0&2&0&\ldots&0\\
{}1&0&1&\ddots&\vdots\\
{}0&1&\ddots&\ddots&0\\
{}\vdots&\ddots&\ddots&0&1\\
{}0&\ldots&0&1&0
\end {array} \right]\,.
 \label{Tsqwodel}
 \ee
Although the latter Hamiltonian matrix only marginally (i.e., in the
single matrix element) differs from its square-well predecessor
(\ref{sqwodel}), both of these models lead to the perceivably
different (though, in both cases, robustly real) spectra of
energies, with Eq.~(\ref{eigvalU}) complemented by the equally
elementary formula
 \be
  E^{[T]}_n=2\,\cos
 \frac{(n+1/2)\pi}{N}\,, \ \ \ \ \ n = 0, 1, \ldots, N-1\,,
 \label{eigvalT}
 \ee
This conclusion may easily be verified by the insertion of the
$N-$dimensional ket-vector ansatz (\ref{eigvecT}) into our
Schr\"{o}dinger Eq.~(\ref{SEwodelT}) + (\ref{Tsqwodel}).

One reveals that our two models share the same formal merit of
having their energies $E_n=2x_n$ given in terms of the roots of the
very similar respective secular equations $U(N,x_n)=0$ and
$T(N,x_n)=0$. Naturally, the manifestly non-symmetric (i.e.,
non-Hermitian) form of the latter Hamiltonian (\ref{Tsqwodel})
violates the analogy and requires an appropriate amendment of the
underlying physical Hilbert space of states (cf. our recent review
\cite{SIGMA} as summarized in Appendix A below). In our present
letter we simply intend to discuss and remove the latter, purely
technical obstacle.

\subsection{The trick of the Hermitization }

Among applications of quantum theory there exists a striking
contrast between the widely employed variability and adaptability of
Hamiltonians $H$ (which, by construction, have to generate the
unitary time evolution) and a very rare use of the equally
admissible variability of the eligible (i.e., symbolically,
$\kappa-$numbered or $\vec{\kappa}-$numbered) alternative
representations of the Hilbert space ${\cal H}={\cal
H}(\vec{\kappa}_0)$ of admissible states.

In our research we felt addressed by this disproportion. Naturally,
we understood that the widespread practice of choosing,
predominantly, just the ``most friendly" Hilbert-space
representation ${\cal H}^{(F)}= L^2(\mathbb{R})$ (and/or its
inessential modifications) is only based on certain extremely formal
grounds (cf. the critique of this purely comfort-based habit, say,
in \cite{Carl}).

In the notation summarized briefly in Appendix A below let us
consider, therefore, the less conventional and less restrictive
approach to quantum model-building in which a given quantum system
is described not only by its Hamiltonian $H$ (which will be,
generically, non-Hermitian in the randomly selected ``first-option"
Hilbert space ${\cal H}^{(F)}$) but also by the related (and,
generically, necessarily nontrivial) suitable metric operator
$\Theta \neq I$.

In this context our present paper intends to offer a sufficiently
transparent illustration of the feasibility of working with variable
multiindices $\vec{\kappa}$ in the operator doublets
$(H,\Theta(\vec{\kappa}))$. For this purpose we shall accept a few
drastic simplifications of the technicalities, assuming that

\begin{itemize}

\item
both $H$ and $\Theta$ are just the real, $N$ by $N$-dimensional
matrices with an arbitrary finite (though freely variable) dimension
$N=1,2,\ldots$;

\item
the ``well-studied" consequences of the possible variations of $H$
will be ignored and suppressed; just the single sample of $H$ will
be considered.

\end{itemize}

 \noindent
We shall mainly be interested in the physics-influencing
consequences of the variability of $\Theta=\Theta(\vec{\kappa})$.
Once we intend to study just nontrivial metrics, the Hamiltonian
matrix itself will be required non-Hermitian in the trivial (and,
hence, unphysical) Hilbert space ${\cal H}^{(F)}$, $H \neq
H^\dagger$.

\subsection{The nearest-neighbor-interaction lattices}

Among all of the finite-dimensional Schr\"{o}dinger equations
 \be
 \hat{H}\, |\psi^{(N)}_n\kt= E^{(N)}_n\,
|\psi^{(N)}_n\kt\,, \ \ \ \ \ n = 0, 1, \ldots, N-1\,
 \label{SEwodel}
  \ee
with the $N$ by $N$ Hamiltonian matrices $\hat{H}$ possessing real
spectra $\{E^{(N)}_n\}$ a privileged, most efficiently solvable
subclass is formed by the ``chain" models in which the matrix $H$ is
tridagonal,
 \be
 \hat{H}=
  \left[ \begin {array}{cccccc}
   a_1&c_1&0&0&\ldots&0
  \\
  b_2
 &a_2&c_2&0&\ldots&0
 \\0&b_3&a_3&c_3&\ddots&\vdots
 \\0&0
 &\ddots&\ddots&\ddots&0
 %\\
 % {}\vdots&\ddots&\ddots
 %&b_{N-2}&a_{N-2}&c_{N-2}&0
 \\\vdots&\ddots&\ddots&b_{N-1}&a_{N-1}&c_{N-1}
 \\{}0&\ldots&0&0&b_{N}&a_{N}\\
 \end {array} \right]\,.
 \label{kitiely}
 \ee
Often, these models are studied as mimicking an $N-$site
quantum-lattice dynamics containing just the nearest-neighbor
interaction (cf., e.g., \cite{chain}).

A simplification of the model may be achieved when our
Schr\"{o}dinger Eq.~(\ref{SEwodel}) + (\ref{kitiely}) becomes
solvable in closed form. Two illustrations of the specific merits of
such an approach may be found in our recent
papers~\cite{gegenb,lagenre}. We studied there the exactly solvable
models in which the {\em explicit} form of the eigenvectors
$|\psi^{(N)}_n\kt$ was obtained in the following two alternative
forms defined either in terms of the Gegenbauer classical orthogonal
polynomials $G(n, a, x)$,
 \be
 |\psi^{(N)}_n\kt\,\equiv\,|\psi_n^{[G]}(a)\kt=\left[
 \begin {array}{c}
 G(0, a, E_n)\\
 G(1, a, E_n)\\
 \vdots\\
 G(N-1, a, E_n)
 \end {array} \right]\,
 \label{eigvecG}
 \ee
or in terms of the Laguerre polynomials $L(n, a, x)$,
 \be
 |\psi^{(N)}_n\kt\,\equiv\,|\psi_n^{[L]}(a)\kt=\left[
 \begin {array}{c}
 L(0, a, E_n)\\
 L(1, a, E_n)\\
 \vdots\\
 L(N-1, a, E_n)
 \end {array} \right]\,.
 \label{eigvecL}
 \ee
Due to the necessity of having a Hamiltonian with a free parameter
$\lambda\equiv a$ at our disposal, we had to accept the serious
shortcoming of both of these models, viz, the purely numerical
character of the determination of the respective energies
$E^{(N)}_n\,\equiv\,E^{[G]}_n(a)$ and
$E^{(N)}_n\,\equiv\,E^{[L]}_n(a)$ at the general dimension $N$.

In this context our present paper has been motivated by the
intention to simplify the calculations and, in particular, to get
rid of the necessity of the numerical determination of the energies.

\section{The variability of the metric for $H=H^{[T]}$
\label{III}}

%
%
%\subsection{The alternative, apparently non-Hermitian Chebyshev
%quantum lattice  }

Recalling the experience gained within the ${\cal PT}-$symmetric
quantum mechanics one may employ a ``sophisticated" physical Hilbert
space with nontrivial metric, viz., ${\cal H}^{(S)}\neq {\cal
H}^{(F)}$. Then, the Hermiticity of the matrix $H$ in simple-minded
space ${\cal H}^{(F)}$ need not be considered obligatory
\cite{Carl}. In this setting our present paper may be perceived as
inspired by Refs.~\cite{gegenb,lagenre}, i.e., by the feasibility of
the construction of the solvable chain models of the form
(\ref{eigvecG}) and (\ref{eigvecL}). The non-numerical nature of the
related constructions of the metrics enabled us to obtain the fairly
large, multi-labeled set of the user-friendly ``Hermitization"
recipes which varied with  a multiindex $\vec{\kappa}$ in metrics
$\Theta=\Theta(\vec{\kappa})$.

In a short detour let us mention that the multiindex-controlled
variability of the metrics has been first considered useful in
nuclear physics \cite{Geyer}. The implementation of the fully
general description of a quantum system by means of the preselected
pair of operators $H(\lambda)$ {\em and} $\Theta(\kappa)$ has been
recommended there as a direct tool of the construction of
phenomenological models. The different choices of the metric have
been identified there with the alternative specifications of the
different admissible sets of operators of observables. In this
sense, the generic change of multiindex $\lambda$ or $\kappa$ plays
the same dynamical role of the control of measurable predictions.

The domain ${\cal D}^{(H)}$ of the variability of the
Hamiltonian-controlling or ``coupling" parameters $\lambda$
themselves would be bounded by the points at which the spectrum of
$H(\lambda)$ (i.e., of energies of the system) ceases to be real
and, hence, observable. These points of boundary $\partial {\cal
D}^{(H)}$ are called exceptional points. Many years ago their exact
definition has been given by Kato \cite{Kato}. At present, their
phenomenological  role is being intensively studied \cite{FP}.

The physical interpretation as well as the boundaries of
admissibility specify the second parametric domain ${\cal
D}^{(\Theta)}$  of the physics-compatible set of parameters $\kappa$
(entering the metric) in a less straightforward manner. This
inspired our present study. We shall select a parameter-free
Hamiltonian $H$ and vary just the metrics $\Theta=\Theta(\kappa)$.
In this manner we shall be able to study the mechanisms by which the
concept of the horizon of the observability of the underlying
quantum system is put into a Hamiltonian-independent perspective.

\subsection{The closed formula for the metrics
$\Theta=\Theta^{[T]}(\vec{\kappa})$.}

Given the two alternative initial $N$ by $N$ tridiagonal real
matrices (\ref{sqwodel}) and (\ref{Tsqwodel}) one would have a
tendency of skipping,  as unphysical, the latter possibility with
non-Hermitian Hamiltonian $H^{[T]}\neq \left (H^{[T]}\right
)^\dagger$. In the spirit of the conventional textbooks only the
former option with $H^{[U]}= \left (H^{[U]}\right )^\dagger$ would
survive. In our present, methodically oriented paper we shall follow
the guidance given by Refs.~\cite{gegenb,lagenre} which allows us to
ignore, on the contrary, the former Hermitian Hamiltonian $H^{[U]}$
as not too interesting.

Under the assumption $H \neq H^\dagger$, naturally \cite{Ali},  one
has to replace the single Schr\"{o}dinger Eq.~(\ref{SEwodel}) by the
doublet of the left-eigenvector and right-eigenvector problems.
Using our notation of review paper~\cite{SIGMA} with doubly-marked
left eigenvectors this pair of equations reads
 \be
 H\,|\psi_n\kt = E_n\,|\psi_n\kt\,,\ \ \ \ \ \
 \bbr \psi_n|\,H=E_n\,\bbr \psi_n|\,
 \ee
or, equivalently,
 \be
 H\,|\psi_n\kt = E_n\,|\psi_n\kt\,,\ \ \ \ \ \
  H^\dagger\,|\psi_n\kkt = E_n^*\,|\psi_n\kkt\,.
 \ee
Under the assumption that the spectrum is real and non-degenerate it
is easy to recall the standard rules of linear algebra and to deduce
that the resulting two $N-$plets of eigenvectors form a bicomplete
biorthogonal basis,
 \be
 I = \sum_{n=0}^{N-1}\,|\psi_n\kt\,
 \frac{1}{\bbr \psi_n|\psi_n\kt}\,\bbr \psi_n|\,,
 \ \ \ \ \
 \bbr \psi_m|\psi_n\kt = \delta_{m,n}\,\bbr \psi_n|\psi_n\kt\,.
 \ee
In Ref.~\cite{SIGMAdva} we explained that such a separation of the
kets from the ketkets implies that the Mostafazadeh's \cite{Ali}
general definition of the metric may be written in the
spectral-representation form
 \be
 \Theta= \sum_{n=0}^{N-1}\,|\psi_n\kkt\,
 |\nu_n|^2\,\bbr \psi_n|\,
 \label{metric}
 \ee
where the $N-$plet of complex parameters $\nu_n \neq 0$ is
arbitrary.

\subsection{The simplified metrics and pseudometrics}

For our present concrete toy-model Hamiltonian (\ref{Tsqwodel}) the
known form (\ref{eigvecT}) of the quantum-lattice-site
ket-components $\{\alpha|\psi^{[T]}\kt$ with $\alpha=1,2,\ldots,N$,
i.e., formulae
 \ben
 \{ 1|\psi^{[T]}\kt=T(0,x)= 1\,,\ \ \  \{ 2|\,\psi^{[T]}\kt=T(1,x)= x
 \,,\ \ \
 \een
 \be
  \{ 3|\psi^{[T]}\kt=T(2,x)=
 2\,{x}^{2}-1
 \,,\ \ \
 \,,\ldots\,, \{ N|\psi^{[T]}\kt=T(N-1,x)
 \label{setTa}
 \ee
may be complemented, after some algebra, by their ketket-component
counterparts
 \be
 \{ 1|\psi^{[T]}\kkt=T(0,x)/2= 1/2\,,\ \ \  \{ \alpha|\,\psi^{[T]}\kkt=T(n,x)
 \,,\ \ \ \alpha= 2, 3,\ldots, N\,.
 \label{setTb}
 \ee
This means that the difference between these two sets only emerges
at the single site-index $\alpha=1$. As the main  consequence of
such a not entirely expected simplicity of the left eigenvectors
(\ref{setTb}), the above-mentioned general $N-$term
spectral-expansion sum (\ref{metric}) + (\ref{setTb}) may be
reclassified, for our $H=H^{[T]}$, as a very rare closed-form
definition of the complete set of the matrices of the eligible
metrics.

Naturally, the latter formula covers all of the possible special
cases and choices of the optional $n-$plets of parameters $\nu_n
\neq 0$. In this sense its practical usefulness is universal. Still,
the resulting generic form of the metric is a general matrix
containing as many as $N$ free parameters $|\nu_n|^2>0$.
Fortunately, in the light of the experience gained in
Refs.~\cite{gegenb,lagenre} one can expect that such a complete set
of the eligible metrics can be split into subsets of certain much
simpler, sparse-matrix form.

The same possibility also emerges in our present Chebyshevian
quantum lattice. The sparsity of the metrics may be achieved when
one replaces the fully general recipe (\ref{metric}) + (\ref{setTb})
by its implicit, linear-algebraic-equation  version in which a
suitable ansatz for $\Theta$ is inserted in matrix equation
 \be
 H^\dagger \Theta = \Theta\,H\,.
 \label{dieudo}
 \ee
Surprisingly enough, such an approach will enable us to make full
use of the chain-model tridiagonality (\ref{kitiely}) of $H$ and to
complement it by the tentative parallel assumption of the
band-matrix form of the metrics.

Once this idea is implemented, one immediately arrives at the
simplest possible (and, up to an irrelevant overall factor,
parameter-independent) form of the diagonal-metric solution of
Eq.~(\ref{dieudo}),
 \be
 \Theta^{(diagonal)}_{\alpha,\beta}=
 \delta_{\alpha,\beta}(1-\delta_{\alpha,1}/2)
 \,\Theta^{(diagonal)}_{N,N}\,,
 \ \ \ \ \ \alpha,\beta = 1,2, \ldots,N\, .
 \ee
By the similar technique, the next, one-parametric tridiagonal
candidate for the metric is found to have the universal
$N-$dimensional matrix form
 \be
 \Theta^{(N)}({\lambda})=
 \left[ \begin {array}{cccccc}
 1/2&{\lambda}&0&0&\ldots&0\\
 {}{\lambda}&1&{\lambda}&0 &\ddots&\vdots\\
 {}0&{\lambda}&1&{\ddots}&\ddots&0\\
 {}0&0&{\ddots}&\ddots&{\lambda}&0 \\
 {}\vdots&\ddots&\ddots&{\lambda}&1&{\lambda}\\
 {}0&\ldots&0&0&{\lambda}&1
 \end {array} \right]\,.
 \label{Ka}
 \ee
Next, the pentadiagonal, two-parametric formula is obtained,
 \be
 \Theta^{(N)}({\lambda,\mu})=
 \left[ \begin {array}{cc|cccc|c}
 1/2&{\lambda}&\mu&0&\ldots&0&0\\
 {}{\lambda}&1+\mu&{\lambda}&\mu &0&\ldots&0\\
 \hline
 \mu&{\lambda}&1&{\lambda}&\mu&\ddots&\vdots\\
 0&\mu&{\lambda}&1&{\ddots}&\ddots&0\\
 {}0&0&\mu&{\ddots}&\ddots&{\lambda}&\mu \\
 {}\vdots&\vdots&\ddots&\ddots&{\lambda}&1&{\lambda}\\
 \hline
 {}0&0&\ldots&0&\mu&{\lambda}&1-\mu
 \end {array} \right]\,,
 \label{Kabel}
 \ee
with the only irregularities at the triplet of diagonal elements
$\Theta^{(N)}_{j,k}({\lambda,\mu})$ with $(j,k)= (1,1), (2,2) $ and
$(N,N)$. Etc.

\section{The  positivity of the metrics
 \label{IV}}

By the universal, non-sparse-matrix formula (\ref{metric}) the
necessary condition of the positivity of the metric is trivially
satisfied. Once we restrict our attention to the sparse-matrix
metrics, the problem of the guarantee of the positivity of the
solution $\Theta$ of Eq.~(\ref{dieudo}) reemerges and has to be
treated, e.g., perturbatively \cite{lagenre}. For our present,
extremely simple toy model $H^{[T]}$, a few much stronger results
(or at least conjectures) can be found and formulated, indeed.

\subsection{The eigenvalues of matrices  $\Theta(\vec{\kappa})$}

%domains
%${\cal D}^{(\Theta)}$  of the Hermiticity of $H^{[T]}$}

The price to be paid for the sparsity of the illustrative matrices
(\ref{Ka}) and (\ref{Kabel}) (which were obtained via an appropriate
ansatz and direct brute-force solution of the set of $N^2$ linear
algebraic Eqs.~(\ref{dieudo})) lies in the loss of connection with
Eq.~(\ref{metric}), i.e., in the loss of the direct control of the
necessary positivity of  {\em all} of the free parameters
$|\nu_n|^2$. Fortunately, such a shortcoming may be perceived as
more than compensated not only by the almost complete survival of
the mathematical simplicity of the formulae for the inner products
based on nontrivial $\Theta\neq I$ (cf. Appendix A) but, first of
all, by the parallel partial preservation of the physical concept of
the (possibly, ``smeared" ~\cite{lagenre}) locality of the
interaction (cf. also a few comments in Refs.~\cite{Jones} and
\cite{cubic} in this respect).

In opposite direction, it is easily verified by immediate insertions
that the direct use of formula (\ref{metric}) + (\ref{setTb}) always
determines just the generic metric in its generic form of a full,
``long-ranged" matrix. This is the ``worst possible" scenario, from
the point of physics at least. Indeed, in such a case our initial,
intuitive but rather natural ``nearest-neighbor-interaction"
interpretation of Hamiltonians (\ref{kitiely}) is lost.

A reasonable (i.e., recommended) compromise may be seen in the use
of the few-parametric and band-matrix forms of the metrics which
remain easily obtainable and which are sampled by Eqs.~(\ref{Ka})
and (\ref{Kabel}). This class of solutions of Eq.~(\ref{dieudo}) may
be perceived as mathematically exceptional as well as intuitively
appealing. In phenomenological context, they will certainly
facilitate any physics-determining specification of the
metric-compatible forms of all of the eligible operators
representing, in principle, the observable quantities \cite{Geyer}.

A more quantitative argument which might support the preference of
the band-matrix metrics $\Theta(\vec{\kappa})$ emerged during the
purely numerical study of certain cryptohermitian discrete quantum
graphs \cite{submIII}. Empirically we found out, typically, that the
volume of the domain of the positivity of the metrics (to be denoted
here by the symbol ${\cal D}^{(\Theta)}$) did not get too small when
one restricted attention just to a sparse, band-matrix form of the
metric.

\subsection{The horizons of the observability  $\partial {\cal
D}^{(\Theta)}$ at $N=6$}

In Eq.~(\ref{Ka}) and the like we need not necessarily be willing or
able to guarantee that the parameter $\lambda$ lies in the
corresponding physical domain ${\cal D}^{(\Theta)}$. In such a case
we shall sometimes denote the matrix $\Theta^{(N)}(\lambda)$ by the
symbol $K^{(N)}(\lambda)$ (in tridiagonal case) or
$L^{(N)}(\lambda)$ (in pentadiagonal case) etc. Such matrices  may
still play the role of Krein-space or Pontryagin-space
pseudometrics, provided only that none of their eigenvalues vanishes
\cite{Langer}.

\begin{figure}[h]                     %instead of \begin{figure}[t]
\begin{center}                         %instead of \begin{center}
\epsfig{file=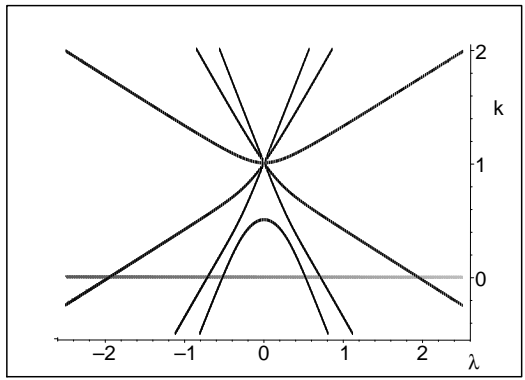,angle=270,width=0.5\textwidth}
\end{center}                         %instead of \end{center}
\vspace{-2mm}\caption{The $\lambda-$dependence of the sextuplet of
the eigenvalues of matrix $K^{(6)}(\lambda)$ of Eq.~(\ref{Ka}).
 \label{fi0}}
\end{figure}

The first graphical guide to the discussion of these characteristics
of the formal solution $K^{(N)}(\lambda)$ of Eq.~(\ref{dieudo}) is
provided, at $N=6$, by Fig.~\ref{fi0} where all of the six
eigenvalues $k_j$ of $K^{(N)}(\lambda)$ are displayed as functions
of an unrestrictedly varied optional parameter $\lambda$. This
illustrative graph might be replaced by the easy rigorous proofs
(left to the readers) that

(1) matrix $K^{(6)}(\lambda)$ defines the (positive definite) metric
$\Theta^{(6)}(\lambda)$ if and only if $ |\lambda| < 0.5176380902=2
\lambda^{(6)}_{min}$ where $\lambda^{(6)}_{min}=0.2588190451$  is
the smallest positive zero of $T(6,\lambda)$;

(2) matrix $K^{(6)}(\lambda)$ specifies the parity-resembling
pseudometric ${\cal P}^{(6)}(\lambda)$ (with the three positive and
three negative eigenvalues) if and only if $ |\lambda|
> 1.931851653=2 \lambda^{(6)}_{max}$ where
$\lambda^{(6)}_{max}=0.9659258265$  is the largest zero  of
$T(6,\lambda)$;

(3) finally, matrix $K^{(6)}(\lambda)$ possesses strictly one
negative eigenvalue for $ 2\,\lambda^{(6)}_{min}< |\lambda| <
\lambda^{(6)}_{med}$ where $\lambda^{(6)}_{med}=0.7071067812$  is
the remaining, third positive zero of $T(6,\lambda)$.

%\newpage
%********** Figure 0 zde

At the sufficiently small real values of parameter $\lambda$ our
matrices $K^{(N)}(\lambda)$ remain positive definite, i.e., eligible
as the desired and necessary metrics at any $N$. All of them define
the corresponding particular admissible physical Hilbert space.

At $N=6$ it is still easy to move beyond the perturbation regime. It
is possible to prove that up to the single remote and weakly
$\lambda-$dependent lowest eigenvalue $k_1(\lambda) \sim 1/2$ of
$K^{(6)}(\lambda)$, the remaining quintuplet forms a locally linear
left-right symmetric star with the center at the quintuply
degenerate eigenvalue $1$ at $\lambda=0$. Thus, we may set
$k_j(\lambda) = 1+\lambda\,y(\lambda)$ yielding the simplified
secular equation
 $$
 -2\,{\lambda}^{5}{y}^{3}-5\,{\lambda}^{6}{y}^{4}
 +3/2\,{\lambda}^{5}y+6\,{\lambda}^{6}{y}^{2}+1
 /2\,{y}^{5}{\lambda}^{5}+{y}^{6}{\lambda}^{6}-{\lambda}^{6}=0\,.
 $$
Once we factor out $\lambda^5$ we obtain the five leading-order
solutions $y_0(\lambda) \to  0$ and $y^2_{\pm 1}(\lambda)\to 1$ or
$y^2_{\pm 2}(\lambda) \to 3$ in the limit $\lambda \to 0$.

%=======

Another star-shaped dependence of the same eigenvalues emerges at
the large $|\lambda \gg 1$, i.e., asymptotically. At the even $N=2p$
we get strictly $p$ linear asymptotes moving up and strictly $p$
asymptotes moving down with the growth of $|\lambda|$, i.e., a
Krein-space-pseudometric scenario.

At $p=3$ it is still easy to show, non-numerically, that one gets
the sextuplet of asymptotic coefficients $y \approx \pm 1.246979604,
\pm 0.4450418679$ and $\pm 1.801937736$ which, incidentally,
coincide with the six roots of $U(6,y)$. Also a generalization of
this observation to all $p$ would be straightforward.

On this basis one may arrive at a remarkable conclusion that the
variation of $\lambda$ in our matrix $K^{(6)}(\lambda)$ in fact
interpolates between the interval of small $\lambda$ (in which
$K^{(6)}(\lambda)$ may be treated as a metric in an appropriate
Hilbert space) {\em and} the domain of very large $\lambda$ (in
which the same matrix $K^{(6)}(\lambda)$ acquires the meaning of the
metric in Krein space).

%
%
%char polyn for metric N=6:
% $$
%{{\it ben}}^{6}-11/2\,{{\it ben}}^{5}+ \left( {\frac
%{25}{2}}-5\,{x}^{ 2} \right) {{\it ben}}^{4}+ \left( 18\,{x}^{2}-15
%\right) {{\it ben}}^ {3}+ \left( 6\,{x}^{4}-24\,{x}^{2}+10 \right)
%{{\it ben}}^{2}+ \left( -21/2\,{x}^{4}+14\,{x}^{2}-7/2 \right) {\it
%ben}-{x}^{6}+9/2\,{x}^{4}- 3\,{x}^{2}+1/2
% $$
%
%=========

\subsection{The horizons of the observability  $\partial {\cal
D}^{(\Theta)}$ at the larger even $N=2p$}

%ibid for all N - include ex. with N=8:

\begin{figure}[h]                     %instead of \begin{figure}[t]
\begin{center}                         %instead of \begin{center}
\epsfig{file=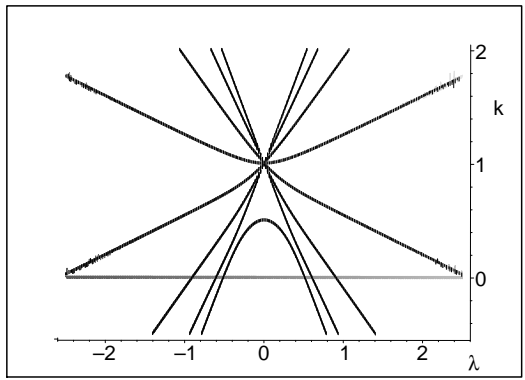,angle=270,width=0.5\textwidth}
\end{center}                         %instead of \end{center}
\vspace{-2mm}\caption{The $\lambda-$dependence of the eigenvalues of
matrix $K^{(N)}(\lambda)$ of Eq.~(\ref{Ka}) at $N=8$.
 \label{fi1}}
\end{figure}

For the sake of brevity let us now skip the cases with the odd
dimensions $N$. Then, taking any even $N=2p$ one specifies, in the
similar manner as above, the domain ${\cal D}^{(\Theta)}$ of the
parameters guaranteeing the positivity of the tridiagonal
band-matrix candidate (\ref{Ka}) for the metric $\Theta$. The
changes of the overall pattern of this construction with $N$ are
inessential. In particular, the above picture may be paralleled by
its $N=8$ descendant displayed in Fig.~\ref{fi1}.

One can summarize that in the tridiagonal case the single-parametric
candidates $\Theta(\lambda)$ become tractable as the true and
positive matrices of the metric, provided only that the selected
value of $\lambda$ lies inside the well defined and $N-$dependent
interval ${\cal D}^{(\Theta)}$. It's boundary points may be shown to
be defined as the, in absolute value, smallest roots of
$U(2p,\lambda)$, i.e.,
 \be
 \lambda \in {\cal D}^{(\Theta)}=\left (\cos
 \frac{(p+1)\pi}{2p+1},\cos
 \frac{(p)\pi}{2p+1} \right )\,, \ \ \ \ \ N=2p\,.
 \ee
From the phenomenological point of view these boundary points may be
perceived as certain ``horizons" of the observability of the system
in question (cf. \cite{SIGMA} for more details).

%\newpage
%********** Figure 1 zde

\subsection{The case of pentadiagonal metrics}

\begin{figure}[h]                     %instead of \begin{figure}[t]
\begin{center}                         %instead of \begin{center}
\epsfig{file=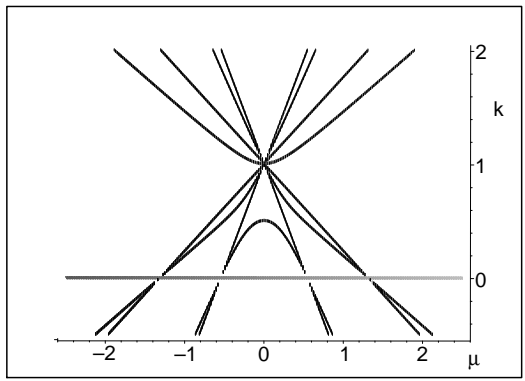,angle=270,width=0.5\textwidth}
\end{center}                         %instead of \end{center}
\vspace{-2mm}\caption{The $\mu-$dependence of the $\lambda=0$
eigenvalues of matrix $L^{(8)}(\lambda,\mu)$.
 \label{fi2}}
\end{figure}

Once we replace the above mentioned tridiagonal ansatz for $\Theta$
by its next, pentadiagonal analogue, you may insert this ansatz in
Eq.~(\ref{dieudo}) and deduce the  general form (\ref{Kabel}) of the
candidate matrix $L^{(8)}(\lambda,\mu)$. Once we set here
$\lambda=0$ we may study the differences from the previous case by
its comparison with Fig.~\ref{fi2}.

%\ne\mupage
%********** Figure 2 zde

A new phenomenon emerges in the form of the degeneracy of vanishing
eigenvalues. In Fig.~\ref{fi2} this may be seen as occurring at
$\lambda=0$ and at $\mu =\sqrt{1 \pm
\sqrt{1/2}}=0.5411961001,1.306562965$.

%\ne\mupage
%********** Figure 3 zde
\begin{figure}[h]                     %instead of \begin{figure}[t]
\begin{center}                         %instead of \begin{center}
\epsfig{file=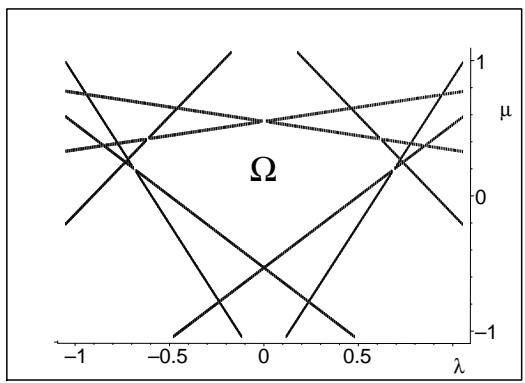,angle=270,width=0.5\textwidth}
\end{center}                         %instead of \end{center}
\vspace{-2mm}\caption{The domain $\Omega$ of positivity of matrix
$L^{(8)}(\lambda,\mu)$.
 \label{fi3}}
\end{figure}

Our last, less restricted and fully two-parametric illustrative
sample of the boundaries of the physical domain for the candidate
matrix $L^{(8)}(\lambda,\mu)$ is provided by Fig.~\ref{fi3}. We see
there that the boundaries (i.e., physical horizons) are formed by
the straight lines. This just reflects the unexpected,
difficult-to-reveal and difficult-to-prove observation (and,
perhaps, an $N-$independent conjecture) that the corresponding
polynomial in two variables which defines the physical horizon
$\partial {\cal D}^{(\Theta)}$ happens to be, in our particular
model, completely factorizable over reals.

\section{Summary \label{V}}

One of the psychologically relevant reasons of the widespread
reluctance of working with the dynamics represented by the metric
(or, more precisely, with the dynamics controlled by the variations
of the set $\kappa$ of the optional parameters in the metric
operator $\Theta=\Theta(\kappa)$) may be seen in the lack of
suitable solvable schematic models. This observation may be
perceived as one of strong motivations of our present proposal. In
the light of Appendix A we may conclude that our manifestly
non-Hermitian toy model (\ref{Tsqwodel}) can really be made
compatible with the standard postulates of quantum theory.

We saw that this goal may be achieved in a multitude of ways. Each
of them will use a particular multiindex $\vec{\kappa} \in {\cal
D}^{\Theta}$ leading to the reinterpretation of the Hamiltonian
which appears manifestly Hermitian in the respective Hilbert space
endowed with the {\em ad hoc} metric $\Theta \neq I$.

We have shown that such a strategy leads to a full theoretical
consistency of our model. One may find it remarkable that even
though our Hamiltonian itself was fixed, the variations of the
metric were still changing its observable physical (i.e., quantum
and probabilistic) contents. The phenomenological core of our
present message may be seen, therefore, in the explicit quantitative
illustration of the changes of the horizons of the validity of the
model which are caused, at the fixed and unique Hamiltonian, by the
mere changes of the parameters in the metric (i.e., by the
variations of the selected physical Hilbert space).

In a formal framework our present model appeared unique by its exact
solvability involving both the energies and both the left and right
eigenvectors. The latter sets form, naturally, a biorthogonal basis
which, unexpectedly, just slightly deviates from its more standard
Chebyshev-polynomial predecessors. One of the other specific strong
merits of the solvability may be seen in the resulting closed-form
prescription which defines {\em all} of the eligible metric
operators $\Theta$ in closed form. In this respect the present model
seems to be entirely unique.

In a complementary approach to the construction of the metrics we
shifted the emphasis from the universal formulae to the alternative
requirement of the sparsity of the metrics. In the light of the
sparsity (tridiagonality) of the Hamiltonian we felt encouraged to
demand also the existence of the matrices $\Theta$ which would be
sparse and less-parametric. Even in this direction we succeeded.
Unfortunately, we were able to indicate, but not able to prove that
there might exist further, not yet revealed aspects of the
solvability of our present model. Preliminarily, this possibility
manifested itself by the piecewise linearity of boundaries of the
physical domain $\Omega \equiv {\cal D}^\Theta$ as sampled in
Fig.~\ref{fi3} at $N=8$.

%\vspace{15mm}

\section*{Acknowledgement}

Work supported by the GA\v{C}R grant Nr. P203/11/1433, by the
M\v{S}MT ``Doppler Institute" project Nr. LC06002 and by the
Institutional Research Plan AV0Z10480505.

%\newpage

%\newpage

\section*{Appendix A. The operators of metric }

It is well known that from the purely mathematical point of view one
has an entirely unrestricted freedom of the choice of the concrete
representation (say, $L^2(\mathbb{R})$) of the abstract Hilbert
space ${\cal H}$ of the admissible states of a quantum system in
question.

In the applications of quantum theory this fact often enables us to
profit from the simplifications provided by the strict unitary
equivalence between various {\em amended} representations obtained,
say, by the Fourier transformation $L^2(\mathbb{R}) \to
L^2(\mathbb{R})$ which converts the so called coordinate
representation (of a single-(quasi)particle system) into the
equivalent momentum representation {\em of the same system}. The
formal reason of this equivalence lies in the unitarity of the
Fourier mapping.

In the same spirit one may replace $L^2(\mathbb{R})$ (i.e., the
Hilbert space of the quadratically integrable functions $f(x)$) by
the slightly more sophisticated Hilbert space $L^2(\mathbb{R},\mu)$
in which one merely replaces the usual condition of the quadratic
integrability of $f(x)$ by the more general integration condition
using a suitable weight $\mu(x)$. The same weighted integration
occurs also in the generalized inner product between $f(x)$ and
$g(x)$ of course,
 \be
 \br f | g\kt :=\int_\mathbb{R}\,f^*(x)g(x) d\mu(x)  \ \ \ \ {\rm in }\ \
 L^2(\mathbb{R},\mu)\,.
 \ee
The next step of generalization has been proposed by Scholtz et al
\cite{Geyer}. They imagined that in the context of nuclear physics
the replacement of the simplest forms of the measure $d \mu(x)$ by
their ``smeared" generalizations might simplify some calculations
while still leaving the general principles of the abstract quantum
theory unchanged. Thus, for the sufficiently smooth measures  $d
\mu(x)= \mu'(x) dx$ this smearing recipe may acquire the
comparatively elementary form of the transition to the double
integration,
 \be
 \int_\mathbb{R}\,f^*(x)g(x) \mu'(x) dx
 \ \ \longrightarrow \ \
 \int_\mathbb{R}\, \int_\mathbb{R}\,f^*(x)\Theta (x,y) g(y) dx\ dy\,.
 \ee
It defines the inner product in the resulting ``smeared" or
``sophisticated" Hilbert-space representation ${\cal H}^{(S)}$ via
the corresponding smearing operator (usually called ``metric"
operator) $\Theta$.

One of the most transparent and elementary illustrative examples has
been constructed by Mostafazadeh \cite{cubic}. He showed that the
special Hilbert space ${\cal H}^{(S)}$ endowed with the {\em
strongly nonlocal} metric of the form
 \be
 \Theta (x,y)\ \sim \ \cosh (\omega)\delta(x-y)
 - \sinh (\omega)\delta(x+y)
 \ee
may be assigned the standard and consistent quantum-mechanical
probabilistic interpretation for {\em every} quantum Hamiltonian $H$
that can be expressed in terms of the single-particle operators
$\hat{x}^2$, $\hat{x}\hat{p}+\hat{p}\hat{x}$ and $\hat{p}^2$ (i.e.,
in particular, for the Hamiltonian $H_0=\hat{p}^2$ of a freely
moving particle).

In fact, the formal background of our present paper will,
incidentally, start from the  discrete representation $\ell^2$ of
the Hilbert space rather than from  $L^2(\mathbb{R})$ or
$L^2(\mathbb{R},\mu)$. In this sense we merely would have to
replace, in this Appendix, the symbols of integrations by
summations. At $N < \infty$ we shall start, in effect, from the
``usual" and ``friendly" Hilbert space ${\cal H}^{(F)} \equiv
\mathbb{R}^N$ of the quantum lattice states\footnote{with the most
common ``scalar" inner product of two vectors
$(\vec{a},\vec{b})=\sum_{\alpha=1}^N a_\alpha b_\alpha$}. Once one
accepts a non-Hermitian matrix $H \neq H^\dagger$ and declares the
latter space ``unphysical", a new, ``standard" $N-$dimensional
Hilbert space ${\cal H}^{(S)}$ (in our present letter, over reals)
must be introduced. In such a manner the metric $\Theta$ becomes an
$N$ by $N$ Hermitian matrix with positive spectrum while the
underlying inner product will read
 \be
 (\vec{a},\vec{b})^{(S)}=\sum_{\alpha=1}^N \sum_{\beta=1}^N
 a_\alpha\,
 \Theta_{\alpha,\beta}\,
 b_\beta\,.
 \ee
One can immediately notice that such a recipe will convert our
present, manifestly non-Hermitian Chebyshevian model
(\ref{Tsqwodel}) into the self-adjoint Hamiltonian.

\end{document}